\documentclass[letterpaper, 10 pt, conference]{ieeeconf}  % Comment this line out
\IEEEoverridecommandlockouts                              % This command is only
\usepackage{url}
\usepackage{verbatim}
\usepackage{amsmath}
\usepackage{amssymb}
\usepackage{rotating}
\usepackage{algorithm}
\usepackage[noend]{algpseudocode}

\usepackage{tabularx}
\usepackage{subfigure}
\usepackage{multirow}
\usepackage{siunitx}

\DeclareMathOperator*{\argmax}{arg\,max}

\allowdisplaybreaks

\title{\LARGE \bf
Empirical Analysis of Fictitious Play for Nash Equilibrium Computation in Multiplayer Games
}

\author{Sam Ganzfried%
\thanks{Ganzfried Research, {\tt\small sam.ganzfried@gmail.com}}%
}

\begin{document}

\maketitle
\thispagestyle{empty}
\pagestyle{empty}

\begin{abstract}
While fictitious play is guaranteed to converge to Nash equilibrium in certain game classes, such as two-player zero-sum games, it is not guaranteed to converge in non-zero-sum and multiplayer games. We show that fictitious play in fact leads to improved Nash equilibrium approximation over a variety of game classes and sizes than (counterfactual) regret minimization, which has recently produced superhuman play for multiplayer poker. We also show that when fictitious play is run several times using random initializations it is able to solve several known challenge problems in which the standard version is known to not converge, including Shapley's classic counterexample. These provide some of the first positive results for fictitious play in these settings, despite the fact that worst-case theoretical results are negative.
\end{abstract}

\section{Introduction}
\label{se:intro}
Approximating Nash equilibrium strategies in multiplayer games is a challenging and important problem. In two-player zero-sum games a Nash equilibrium strategy is guaranteed to win (or tie) in expectation against any opposing strategy by the minimax theorem. In games with more than two players there can be multiple equilibria with different values to the players, and following one has no performance guarantee; however, it was shown that a Nash equilibrium strategy defeated a variety of agents submitted for a class project in a 3-player imperfect-information game, Kuhn poker~\cite{Ganzfried18:Successful}. This demonstrates that Nash equilibrium strategies can be successful in practice despite the fact that they do not have a performance guarantee.

While Nash equilibrium can be computed in polynomial time for two-player zero-sum games, it is PPAD-hard to compute for non-zero-sum and games with 3 or more agents and widely believed that no efficient algorithms exist~\cite{Chen05:Nash,Daskalakis09:Complexity}. Counterfactual regret minimization (CFR) is an iterative self-play procedure that has been proven to converge to Nash equilibrium in two-player zero-sum~\cite{Zinkevich07:Regret}. It has been utilized by superhuman agents in two-player poker, first for limit~\cite{Bowling15:Heads-up} and then for no-limit~\cite{Moravcik17b:DeepStack,Brown17:Superhuman}. It can also be run in non-zero-sum and multiplayer games, but has no significant theoretical guarantees for these settings. It was demonstrated that it does in fact converge to an $\epsilon$-Nash equilibrium (a strategy profile in which no player can gain more than $\epsilon$ by deviating) in three-player Kuhn poker, while it does not converge to equilibrium in the larger game of three-player Leduc hold 'em~\cite{Abou10:Using}. It was subsequently proven that CFR guarantees that no weight is put on iteratively strictly-dominated strategies or actions~\cite{Gibson14:Regret}. While for some small games this guarantee can be very useful (e.g., for two-player Kuhn poker a high fraction of the actions are iteratively-dominated), in most realistic games (such as Texas hold 'em) only a very small fraction of actions are iteratively-dominated, and the guarantee is not useful~\cite{Ganzfried19:Mistakes}. Despite a lack of a significant theoretical guarantee, CFR was combined with subgame solving~\cite{Ganzfried15:Endgame} and depth-limited search~\cite{Brown18:Depth,Hu17:Midgame} to produce a superhuman agent for multiplayer poker~\cite{Brown19:Superhuman}.

Fictitious play (FP) is another iterative algorithm that has been demonstrated to converge to Nash equilibrium in two-player zero-sum games (and in certain other game classes), though not in general for multiplayer or non-zero-sum games~\cite{Brown51:Iterative,Robinson51:Iterative}.  While it is not guaranteed to converge in multiplayer games, it has been proven that if it does converge, then the average of the strategies played throughout the iterations constitute an equilibrium~\cite{Fudenberg98:Theory}. Fictitious play has been successfully applied to approximate Nash equilibrium strategies in a 3-player poker tournament to a small degree of approximation error~\cite{Ganzfried08:Computing,Ganzfried09:Computing}. More recently it has also been used to approximate equilibrium strategies in multiplayer auction~\cite{Rabinovich09:Generalised,Rabinovich13:Computing} and national security scenarios~\cite{Ganzfried20b:Parallel}.

Both counterfactual regret minimization and fictitious play can be integrated with Monte Carlo sampling to enable scalability in large imperfect-information games~\cite{Lanctot09:Monte,Heinrich15:Fictitious}. Recently both approaches have also been integrated with deep reinforcement learning to obtain strong performance in forms of poker~\cite{Heinrich16:Deep,Brown19:Deep}. These two approaches were compared in a simplified version of two-player poker called flop hold 'em poker as well as 2-player limit Texas hold 'em, and Deep Counterfactual Regret Minimization~\cite{Brown19:Deep} was shown to outperform Neural Fictitious Self Play~\cite{Heinrich16:Deep}. It was concluded from these results that ``Fictitious Play has weaker theoretical convergence guarantees than CFR, and in practice converges slower.''~\cite{Brown19:Deep} However, this claim is just made based off the performance of specific integrations of CFR and FP with deep reinforcement learning on two-player zero-sum poker variants. The conclusion does not necessarily generalize to other versions of the algorithms, to other games, or to performance in multiplayer games (where neither approach guarantees convergence). 

In this paper we first compare the performance of the core versions of the counterfactual regret minimization and fictitious play algorithms for a variety of numbers of players and strategies. While counterfactual regret minimization is shown to be superior for two-player zero-sum settings, fictitious play produces significantly closer degree of Nash equilibrium approximation for all other settings. Next, we explore several classic challenge problem instances for which the standard version of fictitious play has been shown to not converge to equilibrium. The most notable counterexample is due to Shapley in 1964~\cite{Shapley64:Some}. Recent work has shown that convergence of fictitious play can be significantly improved by using several initial strategy profiles and selecting the best one~\cite{Ganzfried22:Fictitious}. Building on this work, we show that if we apply fictitious play using multiple random initializations, we are able to in fact obtain convergence to Nash equilibrium for some initializations in these challenge instances. By demonstrating that fictitious play outperforms counterfactual regret minimization (which led to superhuman performance in 6-player poker) in non-zero-sum and multiplayer games, and showing that random initializations allow fictitious play to solve notorious challenge instances, we present significant positive results for the performance of fictitious play in these settings, which are in stark contrast to conventional wisdom.

\section{Fictitious Play and Regret Minimization}
\label{se:algorithms}
A strategic-form game consists of a finite set of players $N = \{1,\ldots,n\}$, a finite set of pure strategies $S_i$ for each player $i$, and a real-valued utility for each player for each strategy vector (aka \emph{strategy profile}), $u_i : \times_i S_i \rightarrow \mathbb{R}$. For simplicity we will assume that all players have the same number of pure strategies, $|S_i| = m$ for all $i$. A two-player game is called \emph{zero sum} if the sum of the payoffs for all strategy profiles equals zero, i.e., $u_1(s_1,s_2) + u_2(s_1, s_2) = 0$ for all $s_1 \in S_1, s_2 \in S_2$. If this sum is always equal to some constant $c$ as opposed to zero then the game is called \emph{constant sum}. Games that are constant sum are strategically equivalent to zero-sum games, and all results from zero-sum games (e.g., minimax theorem and convergence of CFR and FP) also hold for constant-sum games.

A \emph{mixed strategy} $\sigma_i$ for player $i$ is a probability distribution over pure strategies, where $\sigma_i(s_{i'})$ is the probability that player $i$ plays $s_{i'} \in S_i$ under $\sigma_i$. Let $\Sigma_i$ denote the full set of mixed strategies for player $i$. A strategy profile $\sigma^* = (\sigma^*_1,\ldots,\sigma^*_n)$ is a \emph{Nash equilibrium} if $u_i(\sigma^*_i,\sigma^*_{-i}) \geq u_i(\sigma_i, \sigma^*_{-i})$ for all $\sigma_i \in \Sigma_i$ for all $i \in N$, where $\sigma^*_{-i}$ denotes the vector of the components of strategy $\sigma^*$ for all players excluding $i$. It is well known that a Nash equilibrium exists in all finite games~\cite{Nash50:Non-cooperative}. In practice all that we can hope for is convergence of iterative algorithms to an approximation of Nash equilibrium. For a given candidate strategy profile $\sigma^*$, define $\epsilon(\sigma^*) = \max_i \max_{\sigma_i \in \Sigma_i} \left[ u_i(\sigma_i,\sigma^*_{-i}) - u_i(\sigma^*_i, \sigma^*_{-i}) \right]$. The goal is to compute a strategy profile $\sigma^*$ with as small a value of $\epsilon$ as possible (i.e., $\epsilon = 0$ would indicate that $\sigma^*$ comprises an exact Nash equilibrium). We say that a strategy profile $\sigma^*$ with value $\epsilon$ constitutes an \emph{$\epsilon$-equilibrium}. For two-player zero-sum games, there are algorithms with bounds on the value of $\epsilon$ as a function of the number of iterations and game size, and for different variations $\epsilon$ is proven to approach zero in the limit at different worst-case rates (e.g.,~\cite{Gilpin12:First}).

In fictitious play, each player plays a best response to the average strategies of his opponents thus far. Strategies are initialized arbitrarily at $t=1$ (for our experiments we will initialize them to put equal weight on all pure strategies, though later in the paper we will revisit the initialization). Then each player uses the following rule to obtain the average strategy:% at time $t$:
$$\sigma^t _i = \left( 1 - \frac{1}{t} \right) \sigma^{t-1} _i + \frac{1}{t} \sigma'^t _i,$$
where $\sigma'^t _i$ is a best response of player $i$ to the profile $\sigma^{t-1} _{-i}$  of the other players played at time $t-1$.
Thus, the final strategy after $T$ iterations $\sigma^T$ is the average of the strategies played in the individual iterations (while the best response $\sigma'^t_i$ is the strategy actually played at iteration $t$). Full pseudocode for the classic version of fictitious play is given in Algorithm~\ref{al:classic-fp}.

\begin{algorithm}[!ht]
\caption{Classic fictitious play for $n$-player games}
\label{al:classic-fp}
\textbf{Inputs}: Game $G$, initial mixed strategies $\sigma^0_i$ for $i \in N$, number of iterations $T$
\begin{algorithmic}
\For {$t = 1$ to $T$}
\For {$i = 1$ to $n$}
\State $\sigma'^t_i = \argmax_{\sigma_i \in \Sigma_i} u_i(\sigma_i,\sigma^{t-1}_{-i})$ 
\State $\sigma^t_i = \left( 1 - \frac{1}{t+1} \right) \sigma^{t-1}_i + \frac{1}{t+1} \sigma'^t_i$ 
\EndFor
\EndFor
\Return $(\sigma^T_1,\ldots,\sigma^T_n)$
\end{algorithmic}
\end{algorithm}

For the core version of counterfactual regret minimization, we first compute the \emph{regret} for not playing each pure strategy $s_i$ as opposed to following strategy $\sigma^t$. This is defined as
$$r^t(s_i) = u_i(s_i,\sigma^t_{-i}) - u_i(\sigma^t_i, \sigma^t_{-i}).$$
The regret for $s_i$ at iteration $t$ is then defined as $R^t(s_i) = \sum_{j=1}^t r^j(s_i).$
Define $R^t_+(s_i) = \max \{R^t(s_i),0 \}.$
Initial strategies $\sigma^1$ are uniform random, as for fictitious play. Then, in the core version of CFR strategies are selected according to the \emph{regret matching} rule:
$$\sigma^{t+1}(s_i) = \frac{R^t_+(s_i)}{\sum_{s'_i \in S_i} R^t_+(s'_i)}.$$
If the denominator $\sum_{s'_i \in S_i} R^t_+(s'_i)$ equals zero then we assign each pure strategy with equal probability.
As for fictitious play, the average of the strategies played over the $T$ iterations is the final output.

\section{Comparison of Fictitious Play and Regret Minimization for Nash Equilibrium Approximation in Multiplayer Games}
\label{se:experiments-cfr}
We ran experiments with the core versions of fictitious play and counterfactual regret minimization on uniform-random games for a variety of number of players $n$ and number of pure strategies $m$. We consider $n = 2,3,4,5$ and $m = 3,5,10$. For each setting of $n$ and $m$ we randomly generated 10,000 games with all payoffs chosen uniformly random in [0,1]. For each game, we ran both algorithms for 10,000 iterations (except for the largest game $n=5$, $m=10$ we used 1,000 iterations). For $n = 2$ we also experimented on zero-sum games where player 1's payoff is selected uniformly in [0,1] and player 2's payoff equals 1 minus player 1's payoff. For each game, we computed the degree of Nash equilibrium approximation $\epsilon$ by the two algorithms, as well as their difference. We compute the 95\% confidence interval for the $\epsilon$ differences. If statistical significance was not obtained after 10,000 games for given values of $n$ and $m$, then we ran 100,000 games. If statistical significance was still not obtained, then we declared a tie. The results from the experiments appear in Table~\ref{ta:results-random}.

The results indicate that CFR outperformed FP in two out of the three two-player zero-sum cases ($m=5,10$), which agrees with the previously described results in two-player zero-sum games and the claim in prior work that CFR has a stronger theoretical convergence guarantee for two-player zero-sum games~\cite{Brown19:Deep}. However, for all other cases (3--5 player and 2-player general-sum) fictitious play outperformed counterfactual regret minimization (with all but two of the results statistically significant). %FP seemed to outperform CFR by more as the number of players increased, with largest improvement for the 5-player settings. %This indicates that fictitious play would be particularly superior to CFR for settings with larger numbers of agents. 

We next experimented on several games produced from the GAMUT generator~\cite{Nudelman04:Run}. We used the same games as used in prior work comparing algorithms for multiplayer Nash equilibrium~\cite{Berg17:Exclusion}. We used the variants with 5 players and 3 actions per player (though for two classes we were forced to use 5 actions per player). As prior work had done, we used 2 for the number of facilities parameter in the congestion game class, and for the covariant game we used the minimum possible value, which is $r = -0.25$ for 5 players. All other parameters were generated randomly (as the prior experiments had done). We generated 1,000 games from each class using these distributions. For several of the games generated some of the payoffs were ``NaN,'' and we removed the games in which this occurred. Since the magnitudes of the payoffs varied widely between game classes, we normalized all payoffs to be in [0,1] before applying the algorithms (by subtracting the smallest payoff from all the payoffs and then dividing all payoffs by the difference between the max and min payoff). Note that linear transformations of the payoffs exactly preserve Nash equilibria, so this normalization has no effect on the solutions. For these experiments we used 10,000 iterations of both algorithms. The results are given in Table~\ref{ta:results-gamut}. For all game classes other than two-player zero sum, fictitious play outperformed CFR with statistical significance or the performances were statistically indistinguishable. %Both algorithms achieved small values of $\epsilon$ for each of the game classes (in most cases the value was extremely small), despite the fact that they are not guaranteed to converge to Nash equilibrium for multiplayer games.
%For several of the game classes CFR had an extremely high value of $\epsilon$, while FP had a relatively small $\epsilon$ value for all cases, and extremely small for most.

%\renewcommand{\tabcolsep}{3pt}
\begin{table*}
\centering
%\footnotesize
%\small
\begin{tabular}{|*{8}{c|}} \hline
$n$ &$m$ &\# games &\# iterations &Avg. CFR $\epsilon$ &Avg. FP $\epsilon$ &Avg. difference in $\epsilon$ &Winner\\ \hline
2 (zs) &3 &10,000 &10,000 &0.00139 &0.00133 &\num{5.945e-5} $\pm$ \num{9.511e-6} &FP\\ \hline
2 (zs) &5 &10,000 &10,000 &0.00239 &0.00261 &\num{-2.219e-4} $\pm$ \num{1.550e-5} &CFR\\ \hline
2 (zs) &10 &10,000 &10,000 &0.00282 &0.00464 &\num{-0.0018} $\pm$ \num{2.277e-5} &CFR\\ \hline
2 &3 &10,000 &10,000 &\num{8.963e-4} &\num{8.447e-4} &\num{5.155e-5} $\pm$ \num{3.934e-5} &FP\\ \hline
2 &5 &100,000 &10,000 &0.00383 &0.00377 &\num{6.000e-5} $\pm$ \num{5.855e-5} &FP\\ \hline
2 &10 &100,000 &10,000 &0.01249 &0.01244 &\num{4.865e-5} $\pm$ \num{1.590e-4} &Tie\\ \hline
3 &3 &100,000 &10,000 &0.00768 &0.00749 &\num{1.897e-4} $\pm$ \num{1.218e-4} &FP\\ \hline
3 &5 &100,000 &10,000 &0.02312 &0.02244 &\num{6.784e-4} $\pm$ \num{2.454e-4} &FP\\ \hline
3 &10 &10,000 &10,000 &0.05963 &0.05574 &0.0039 $\pm$ 0.0012 &FP\\ \hline
4 &3 &100,000 &10,000 &0.01951 &0.01950 &\num{9.798e-6} $\pm$ \num{2.195e-4} &Tie\\ \hline
4 &5 &10,000 &10,000 &0.05121 &0.04635 &0.0049 $\pm$ 0.0011 &FP\\ \hline
4 &10 &10,000 &10,000 &0.08315 &0.06661 &0.0165 $\pm$ \num{8.910e-4} &FP\\ \hline
5 &3 &10,000 &10,000 &0.03505 &0.03303 &0.0020 $\pm$ \num{8.921e-4} &FP\\ \hline
5 &5 &10,000 &10,000 &0.06631 &0.05447 &0.0118 $\pm$ \num{8.896e-4} &FP\\ \hline
5 &10 &10,000 &1,000 &0.06350 &0.04341 &0.0201 $\pm$ \num{5.509e-4} &FP\\ \hline
\end{tabular}
\caption{Results for uniform-random games. First column is number of players $n$ (and whether zero-sum for $n=2$); second column is number of pure strategies per player $m$; third column is number of games generated; fourth column is number of iterations for CFR and FP; fifth column is average value of $\epsilon$ from CFR; sixth column is average value of $\epsilon$ from FP; seventh column is average value of the $\epsilon$ difference, with the 95\% confidence interval (positive value indicates that FP outperformed CFR).}
\label{ta:results-random}
\end{table*}

\begin{table*}
\centering
%\footnotesize
%\small
\begin{tabular}{|*{7}{c|}} \hline
Game class &$m$ &\# games &Avg. CFR $\epsilon$ &Avg. FP $\epsilon$ &Avg. difference in $\epsilon$ &Winner\\ \hline
Bertrand oligopoly &3 &962 &\num{2.54e-5} &\num{1.714e-5} &\num{8.262e-6} $\pm$ \num{9.575e-7} &FP \\ \hline
Bidirectional LEG  &3 &1000 &\num{4.212e-5} &\num{3.561e-5} &\num{6.513e-6} $\pm$ \num{3.084e-6} &FP \\ \hline
Collaboration &5 &1000 &\num{4.873e-4} &\num{4.428e-4} &\num{4.452e-5} $\pm$ \num{9.104e-5} &Tie \\ \hline
Congestion &3 &1000 &\num{3.066e-5} &\num{2.414e-5} &\num{6.521e-6} $\pm$ \num{2.103e-6} &FP \\ \hline
%Coordination &5 &1000 &0.00120 &\num{3.879e-4} &\num{8.134e-4} $\pm$ \num{4.088e-4} &FP \\ \hline
Covariant &3 &1000 &0.02453 &0.02333 &0.00120$\pm$ 0.00118 &FP \\ \hline
Polymatrix &3 &984 &\num{6.046e-4} &\num{6.289e-4} &\num{-2.434e-5} $\pm$ \num{1.066e-4} &Tie \\ \hline
Random graphical &3 &1000 &0.01788 &0.01582 &0.00206 $\pm$ 0.00201 &FP \\ \hline
Random LEG &3 &1000 &\num{3.628e-4} &\num{4.103e-4} &\num{-4.751e-5} $\pm$ \num{3.910e-4} &Tie \\ \hline
Uniform LEG &3 &1000 &\num{4.382e-5} &\num{2.804e-5} &\num{1.578e-5} $\pm$ \num{1.275e-5} &FP \\ \hline 
\end{tabular}
\caption{Results for 5-player GAMUT games. For both CFR and FP 10,000 iterations were used. The second column is the number of actions available per player (we used 3 when it was available, but some game classes only allowed us to use 5). For each game class we generated 1,000 random games. We ignored the games where some of the payoffs generated were ``NaN,'' and report the number of valid games in the third column. We normalized payoffs in all games to be in [0,1] to make comparison sensible.}
\label{ta:results-gamut}
\end{table*}

\section{Random Initialization Solves Notorious Fictitious Play Counterexamples}
\label{se:shapley}
In 1964 Shapley~\cite{Shapley64:Some} devised a family of games for which fictitious play fails to converge to Nash equilibrium. The games are two player non-zero-sum and depicted in Equation~\ref{eq:shapley}.\footnote{It was previously known that fictitious play converges to Nash equilibrium in two-player zero-sum games, and in two-player non-zero-sum games with 2 strategies per player. Subsequently it had been shown to converge to Nash equilibrium in certain other game classes such as $2 \times N$ two-player games with generic payoffs~\cite{Berger05:Fictitious}.} The payoffs satisfy $a_i > b_i > c_i$ and $\alpha_i > \beta_i > \gamma_i$, for $i = 1,2,3.$ One particular instance from this family that has been singled out~\cite{Wiki22:Fictitious} is given in Equation~\ref{eq:instance}. This game can be viewed as a generalized version of Rock-Paper-Scissors~\cite{Wiki22:Fictitious}. Note that we must permute the first two columns then the last two rows in this game to obtain a game that satisfies Shapley's conditions---the permuted game is given in Equation~\ref{eq:instance-permuted}. Shapley's argument is as follows. He assumes that both players play their first pure strategy in the initial round, and then follow a standard simultaneous version of fictitious play (Algorithm~\ref{al:classic-fp}). He then shows that play will cycle such that the runs of pure strategy profiles increase exponentially, failing convergence to the equilibrium frequencies. Note that all games in Shapley's class contain a unique Nash equilibrium, and for the example we have singled out it selects each pure strategy with equal probability. Shapley states,

\begin{quote}
The argument we have given is independent of the tie-breaking rule. With minor modifications it can also handle the case of alternating moves, as well as the case of nonintegral run lengths. The latter implies that the differential-equation version of FP (see~\cite{Brown51:Iterative}) will also fail to converge to the solution~\cite{Shapley64:Some}.
\end{quote}

\begin{equation}
\begin{bmatrix}
(a_1,\beta_1) & (c_2,\gamma_1) &(b_3,\alpha_1) \\
(b_1,\alpha_2) & (a_2,\beta_2) &(c_3,\gamma_2) \\
(c_1,\gamma_3) & (b_2,\alpha_3) &(a_3,\beta_3) \\
\end{bmatrix}
\label{eq:shapley}
\end{equation}
%\caption{Shapley's family of two-player games.}
%\label{fi:shapley}
%\end{figure}

\begin{equation}
\begin{bmatrix}
(0,0) & (2,1) &(1,2) \\
(1,2) & (0,0) &(2,1) \\
(2,1) & (1,2) &(0,0) \\
\end{bmatrix}
\label{eq:instance}
\end{equation}
 
\begin{equation}
\begin{bmatrix}
(2,1) & (0,0) &(1,2) \\
(1,2) & (2,1) &(0,0) \\
(0,0) & (1,2) &(2,1) \\
\end{bmatrix}
\label{eq:instance-permuted}
\end{equation}
 
We confirm Shapley's finding empirically for the example game by running fictitious play with the given initialization for a large number of iterations. Using $T$ = 100,000 we obtain strategy frequencies of player 1 of $(0.3097,0.5981,0.0922)$, and for player 2 of $(0.7290,0.2348,0.0362),$
which are both clearly very far from the equilibrium frequencies. For $T$ = 1,000,000 we obtain strategy frequencies for player 1 of $(0.3870, 0.0598, 0.5522)$, and for player 2 of $(0.1523, 0.0235, 0.8242).$ As Shapley described, the strategies cycle between exponentially-increasing runs of pure strategy profiles, and convergence is not obtained.

The algorithm also fails to converge for any pure-strategy initializations, and therefore Shapley's selection was not special. In fact, we verify that fictitious play also fails to converge to equilibrium in this game if we initialize all pure strategies to be played with probability $\frac{1}{3}$ for each player, despite the fact that this is the unique Nash equilibrium.\footnote{Note that strict Nash equilibria are absorbing states in fictitious play~\cite{Fudenberg98:Theory}; however in the game we are considering the Nash equilibrium is not strict, which is obvious since only pure-strategy equilibria can be strict.}

Recent work has shown that convergence of fictitious play can be significantly improved by using several initial strategy profiles and selecting the best one~\cite{Ganzfried22:Fictitious}. The general approach is depicted in Algorithm~\ref{al:fp-multi-init}. The best-performing approach used maximin initialization based on solving a nonconvex quadratic program. However just generating the strategies uniformly at random performed almost as well and is computationally simpler, so we will experiment with that approach, which is described in Algorithm~\ref{al:rand-init}.

\begin{algorithm}[!ht]
\caption{Fictitious play with multiple initializations}
\label{al:fp-multi-init}
\textbf{Inputs}: Game $G$, set of $K$ initial mixed strategies $\sigma^0_{k,i}$ for $i \in N$ $k = 1\ldots,K$, number of iterations $T$
\begin{algorithmic}
\State $\epsilon^* = \infty$
\For {$k = 1$ to $K$}
\For {$t = 1$ to $T$}
\For {$i = 1$ to $n$}
\State $\sigma'_{k,i} = \argmax_{\sigma_i \in \Sigma_i} u_i(\sigma_i,\sigma^{t-1}_{k,-i})$ 
\State $\sigma^t_{k,i} = \left( 1 - \frac{1}{t+1} \right) \sigma^{t-1}_{k,i} + \frac{1}{t+1} \sigma'^t_{k,i}$ 
\EndFor
\EndFor
\State $\epsilon_k = \max_i \max_{\sigma_i} \left[ u_i(\sigma_i,\sigma^T_{k,-i}) - u_i(\sigma^T_i, \sigma^T_{k,-i}) \right]$
\If {$\epsilon_k < \epsilon^*$}
\State $\sigma^* = \sigma^T_k$
\State $\epsilon^* = \epsilon_k$
\EndIf
\EndFor
\Return $\sigma^*$
\end{algorithmic}
\end{algorithm}

\begin{algorithm}[!ht]
\caption{Generation of uniform initial strategies}
\label{al:rand-init}
\textbf{Inputs}: Game $G$ with $n$ players and $m$ strategies per player
\begin{algorithmic}
\For {$i = 1$ to $n$}
\State $Z_i = 0$
\For {$j = 1$ to $m$}
\State $u = $ uniform random number in (0,1)
\State $\sigma_i(j) = -\ln (u)$
\State $Z_i = Z_i + \sigma_i(j)$
\EndFor
\For {$j = 1$ to $m$}
\State $\sigma_i(j) = \frac{\sigma_i(j)}{Z_i}$
\EndFor
\EndFor
\Return $\sigma = (\sigma_1,\ldots,\sigma_n)$
\end{algorithmic}
\end{algorithm}

\begin{figure*}[!ht]
\centering
\includegraphics[scale=0.5]{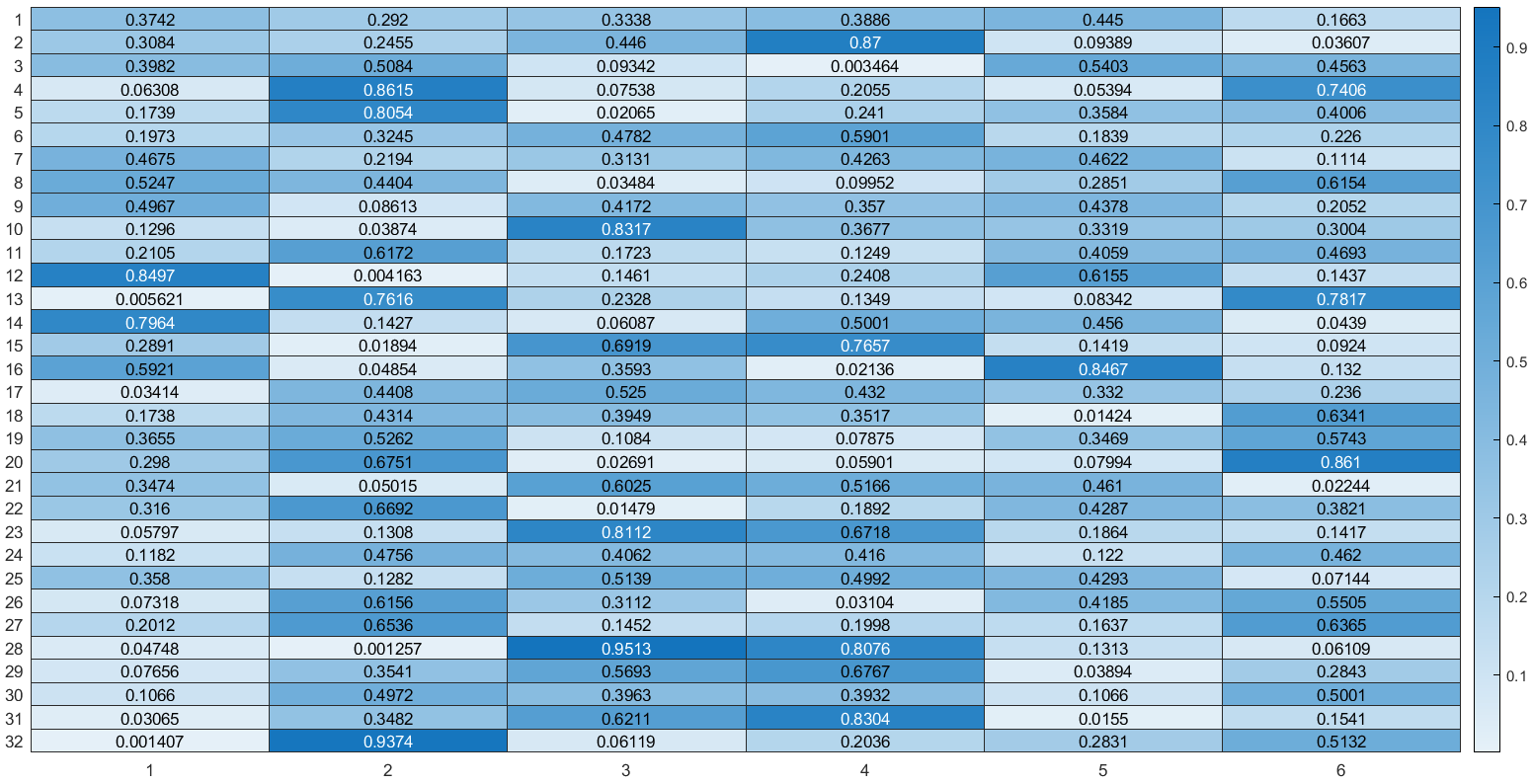}
\caption{Heatmap of the strategy profiles out of the first 100 initializations that produce $\epsilon < \num{e-4}$ in Shapley's game. The first three columns are player 1's strategy probabilities and next three columns are player 2's probabilities.}
\label{fi:heatmap}
\end{figure*}

We applied this approach to solve Shapley's game using 100,000 different initializations generated using Algorithm~\ref{al:rand-init} with $T = 100,000.$
We found that 33,403 of them produced values of $\epsilon$ (i.e., $\epsilon_k$ from Algorithm~\ref{al:fp-multi-init}) smaller than $\num{e-4}.$ So fictitious play converged to the equilibrium for approximately $\frac{1}{3}$ of the initializations. %So in expectation if we run fictitious play using three random initializations, we expect to converge to the equilibrium in one. 
Figure~\ref{fi:heatmap} shows a heatmap of the initializations producing convergence to the Nash equilibrium out of the first 100 initializations. We suspect that the example game is representative and that similar results would apply to the other games in Shapley's family as well.

\begin{figure*}[!ht]
\centering
\includegraphics[scale=0.5]{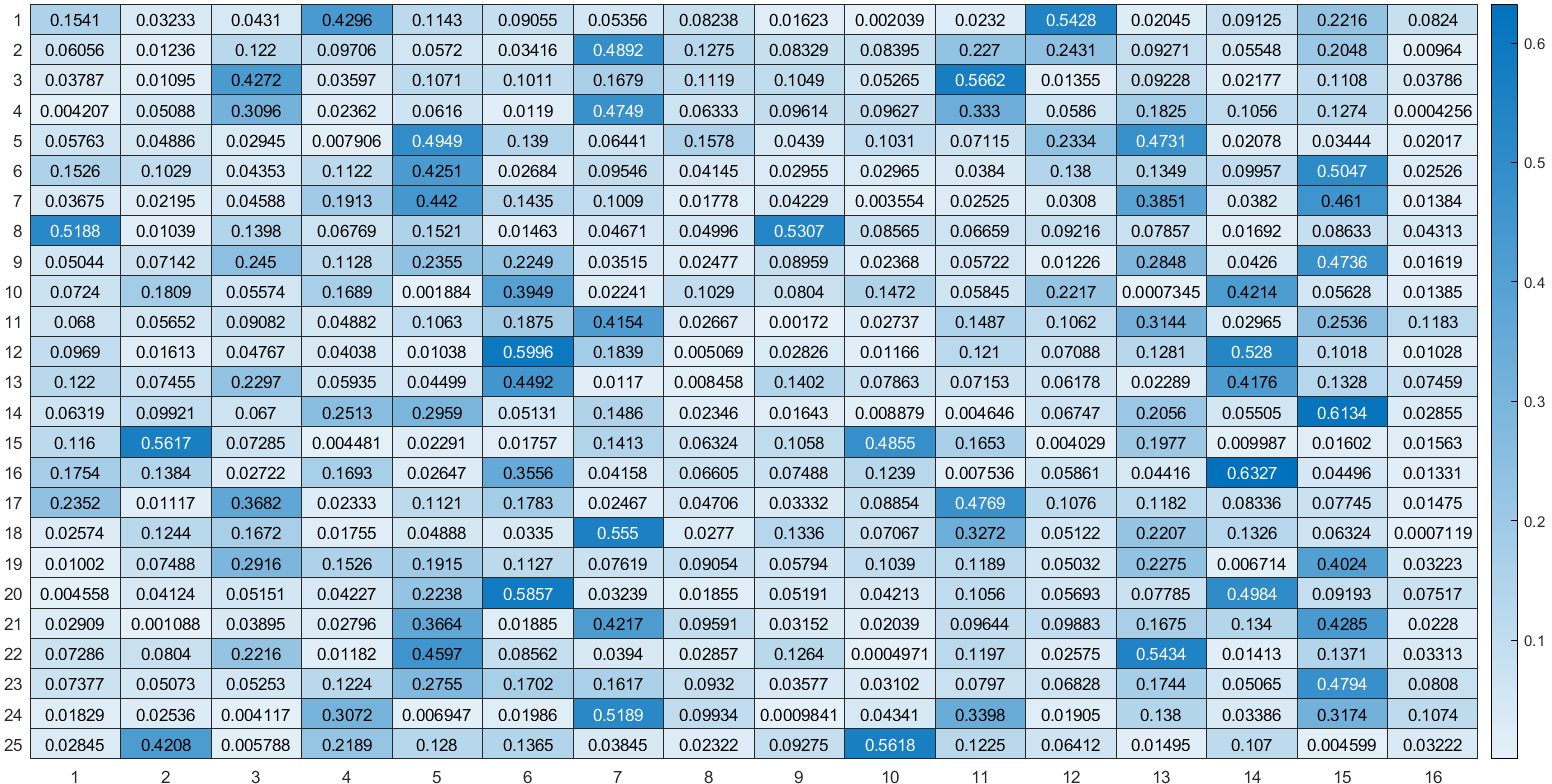}
\caption{Heatmap of the strategy profiles out of the first 10,000 initializations that produce $\epsilon < \num{e-4}$ in Foster and Young's Doctrines Game. The first eight columns are player 1's strategy probabilities and next eight columns are player 2's probabilities.}
\label{fi:heatmap-foster}
\end{figure*}

More recently, Foster and Young have constructed an $8 \times 8$ coordination game in which classic fictitious play does not converge to any of the Nash equilibria~\cite{Foster98:Nonconvergence}. This game is called the Doctrines Game and depicted in Table~\ref{ta:payoffs-doctrines}. We applied Algorithm~\ref{al:rand-init} to this game using 100,000 different initializations with $T = 100,000.$ We observed that 182 of the initializations resulted in values of $\epsilon$ less than $\num{e-4}.$ While this is a relatively small proportion, this nonetheless shows that fictitious play converges to equilibrium for a set of initializations of positive measure. On average, we expect to obtain an $\epsilon$-equilibrium if we run it 550 times. Figure~\ref{fi:heatmap-foster} shows the initializations producing convergence to Nash equilibrium out of the first 10,000.

\renewcommand{\tabcolsep}{1.5pt}
\begin{table}
\centering
%\footnotesize
\small
\begin{tabular}{|*{9}{c|}} \hline
& $A'$ & $A''$ & $B'$ & $B''$ & $C'$ & $C''$ & $D'$ & $D''$\\ \hline
$A'$ &24, 24 &6, 6 &0, 18 &0, 18 &18, 0 &18, 0 &5, 0 &0, 0\\ \hline
$A''$ &6, 6 &24, 24 &0, 18 &0, 18 &18, 0 &18, 0 &4, 0 &0, 0\\ \hline
$B'$ &18, 0 &18, 0 &24, 24 &6, 6 &0, 18 &0, 18 &2, 0 &0, 0\\ \hline
$B''$ &18, 0 &18, 0 &6, 6 &24, 24 &0, 18 &0, 18 &2, 0 &0, 0\\ \hline
$C'$ &0, 18 &0, 18 &18, 0 &18, 0 &24, 24 &6, 6 &1, 0 &0, 0\\ \hline
$C''$ &0, 18 &0, 18 &18, 0 &18, 0 &6, 6 &24, 24 &0, 0 &0, 0\\ \hline
$D'$ &0, 4 &0, 5 &0, 2 &0, 3 &0, 0 &0, 1 &25, 24 &-25, -25\\ \hline
$D''$ &0, 0 &0, 0 &0, 0 &0, 0 &0, 0 &0, 0 &-25, -25 &24, 25\\ \hline
\end{tabular}
\caption{Payoff matrix of the Doctrines Game.}
\label{ta:payoffs-doctrines}
\end{table}

\section{Conclusion}
\label{se:conclusion}
We have shown empirically that fictitious play produces closer degree of Nash equilibrium approximation than regret minimization in two-player non-zero-sum and multiplayer games. This is significant, since counterfactual regret minimization has been demonstrated to obtain superhuman performance in six-player no-limit Texas hold 'em. We have also demonstrated that utilizing multiple random initializations enables fictitious play to solve several classic challenge problems. These results lead to the natural open problem of whether there exist any games for which fictitious play does not converge to Nash equilibrium for any initial strategies (or alternatively, for any space of positive measure). Note that a negative answer to this question would not necessarily violate the P $\neq$ PPAD assumption, since exponentially many initializations may be required or fictitious play may require exponentially many iterations to converge.

%\clearpage
\bibliographystyle{plain} 
\bibliography{C://FromBackup/Research/refs/dairefs}

\end{document}